\documentclass[pdflatex,sn-mathphys]{sn-jnl}
\usepackage{graphicx}
\usepackage{lipsum}
\jyear{2021}
\theoremstyle{thmstyleone}

\theoremstyle{thmstyletwo}

\usepackage{pdfpages} 
\theoremstyle{thmstylethree}

\begin{document}

\title[Article Title]{ 
Quantum echo route towards exceptional points in Anderson localized lasers}
\author[1]{\fnm{Krishna} \sur{Joshi}}

\author*[1]{\fnm{Sushil} \sur{Mujumdar}}\email{mujumdar@tifr.res.in}

\affil*[1]{\orgdiv{Nano-optics and Mesoscopic Optics Laboratory}, \orgname{Tata Institute of Fundamental Research}, \orgaddress{\street{1, Dr. Homi Bhabha Road}, \city{Mumbai}, \postcode{400005}, \state{Maharashtra}, \country{India}}}

\abstract{Exceptional points, that are spectral degeneracies in the parameter space of non-Hermitian systems, have evoked a massive interest in the optical domain owing to their striking consequences on optical behavior of commonly known systems. Through careful engineering of gain and loss, exceptional points have been demonstrated in a variety of photonic systems ranging from optical fibers to chaotic cavities, exhibiting extra-ordinary phenomena and augmented functionalities. However, in the domain of disordered systems, there are still no realizations of exceptional points even though mode-coupling and non-Hermitian behavior is amply demonstrated. The obvious challenge lies in the probabilistic nature of disorder, which is a difficult candidate for parametric control. Here, we exploit the probabilistic nature of Anderson localizing systems by implementing thousands of disorder configurations. We demonstrate statistical occurrences of lasing over exceptional points. Our route towards exceptional points begins with detection of quantum echoes, which are temporal signatures of coupling between modes. Quantum echoes unambiguously set apart two coupled modes from a pair of two isolated modes that are spectrally close perchance. Simultaneous temporal, spectral and spatial investigations provide corroborative evidence of the convergence of eigenvalues and eigenvectors in the approach to the exceptional points. Ultimately, the vanishing of the echo and coalescence of spectral peaks and spatial intensity distributions, accompanied by the square-Lorentzian lineshape of lasing peaks, identify the exceptional point, at which the lasing intensity is seen to be significantly higher.   
}


\maketitle

\section{Main}\label{sec2}

A recent, high-impact development in the field of photonics has been the revelation of parity-time symmetric systems endowed with rich novel physics\cite{bender1999pt, bender1998real, el2018non, ruter2010observation, regensburger2012parity, jing2014pt,lin2011unidirectional,zhu2014p,benisty2011implementation,ding2015coalescence}. At the barycenter of the literature on PT-symmetry in photonics lie peculiar spectral degeneracies called the exceptional points\cite{peng2014parity,peng2014loss, cerjan2016exceptional,feng2013experimental,jing2017high,ding2016emergence,kang2016chiral,zhen2015spawning}. Originating from the realm of quantum mechanics, exceptional points (EPs) signify singularities at which the eigenvalues and eigenfunctions of non-Hermitian matrices cannot be described analytically\cite{heiss2012physics,kato1966analytic}. At the EP, two or more eigenvalues coalesce, simultaneously with the coalescence of associated eigenfunctions\cite{heiss1999phases,moiseyev2011non,el2007theory,pick2017general}. This theoretical construct has led to immense physical consequences in real experimental systems\cite{miri2019exceptional}. Previously discussed in the quantum theory of atomic and molecular resonances\cite{moiseyev1998quantum}, EPs are now demonstrated in carefully designed photonic systems motivated by the upgradation of systemic functionalities, such as, for instance, enhancement in output intensity\cite{feng2014single,hodaei2014parity,takata2021observing} and increased sensing capability\cite{hodaei2017enhanced,chen2017exceptional,hokmabadi2019non}. The ability of precise nanofabrication and on-demand gain/loss has allowed for the realization of several photonic platforms for the observation of EPs\cite{ozdemir2019parity,brandstetter2014reversing}. However, to our knowledge, there are no experimental reports of observations of EPs in disordered mesoscopic systems despite several investigations on non-Hermiticity in such systems\cite{vazquez2014gain,bachelard2022coalescence,davy2019probing,huang2021wave,weidemann2020nonhermitian,balasubrahmaniyam2020necklace,sahoo2022anomalous}.

The procedure of demonstration of exceptional points in photonic structures has typically followed a common theme. The gain or loss parameter in one cavity is systematically tuned while the emission spectrum and the spatial intensity in the sample is monitored. For example, in a coupled microring system, the authors fabricated two physically identical microrings at a pre-defined separation and coupling\cite{hodaei2014parity, hodaei2015parity}.  Optical excitation provided gain, and the loss in one of the cavities was continuously tuned by filtering out the pump energy. At the exceptional point, the lasing exclusively occurred over one microring, and the emission intensity was considerably higher than that from an individual microring pumped at par. This strategy has been adopted in various photonic systems \cite{takata2021observing, kim2016direct, chang2014parity, peng2014loss} that are carefully designed using physically separate cavities with tunable gain/loss. However, such a strategy is not feasible in an Anderson-localizing structure\cite{anderson1958absence,john1987strong,segev2013anderson,chabanov2000statistical,wiersma1997localization,sapienza2010cavity,schwartz2007transport,lahini2008anderson,riboli2011anderson} for several reasons.  Anderson-localizing systems are nondeterministic, wherein the occurrence of a resonance is probabilistic. In a situation of coupled  modes, both the cavities reside in the same physical structure and often involve common scatterers for distributed feedback. Naturally, calibrated introduction of gain/loss into individual localized modes is inconceivable. Hence, the conventional strategy of observing exceptional points is implausible in Anderson-localizing systems. In such a nondeterministic scenario, another route towards an EP can be adopted. The formation of an EP within a localizing system is a probabilistic event. Therefore, if a sufficiently large number of configurations can be generated and monitored, then an exceptional point can be expected with the right probability. Although easily said, this is a challenging prospect because the three ingredients , namely, (i) achieving Anderson localization, (ii) achieving coupled modes thereof, and (iii) addition of adequate gain/loss into the system, are all formidable experimental challenges. Nonetheless, this is the route we adopt in this work, and demonstrate Anderson-localization based lasing over exceptional points. 

We use a technique which allows us to create and monitor thousands of disorder configurations\cite{joshi2020reduction, joshi2022anomalous}. The system is one-dimensional, and hence has an increased probability of Anderson localization and, therefore, coupled Anderson-localized modes.  Optical sources are distributed throughout the structure so that any localized modes existing at any spatial location are readily excited. We employ a simultaneous spectral, spatial and temporal diagnostic setup to completely characterize the emission from the microcavity array. The introduction of temporal diagnostics is a vital step towards identification of coupled modes. Usually, coupling is identified on the basis of spectral splitting. However, the spectra of Anderson localizing systems are random and multimodal in character. One cannot unambiguously distinguish a coupled-mode splitting from a pair of individual isolated modes with comparable spectral separation. On the other hand, the temporal signatures of the two cases are fundamentally different and provide conclusive evidence of splitting. Since temporal diagnostics are hitherto not reported at EP in lasing systems, we first simulate a deterministic prototype lasing structure and introduce the systematic temporal, spectral and spatial behavior at, and during the approach to, an EP. The EP-approach is characterized by quantum echoes (QE) that signify energy transfer between the coupled resonances, equivalent to Rabi oscillations in two-level systems\cite{dietz2007rabi,dembowski2004first}. The echoes vanish at the EP. Subsequently, we illustrate the experimental results in our system wherein the spectra exhibit coupled modes via spectral splitting and the temporal diagnostics simultaneously exhibit quantum echoes. The spatial intensity distributions exhibit the expected supermodes of the coupled system. The approach to an EP is evident in reduced splittings and lower oscillation frequencies, accompanied by increased output intensity. Eventually, emission peaks with largely enhanced intensity are shown, wherein the spatial intensity resides exclusively in one Anderson-localizing cavity. In these cases, the spectral profiles are seen to follow a square-Lorentzian profile, identifying an EP.

\begin{figure}[h!]
\centering
\includegraphics[width=1\textwidth]{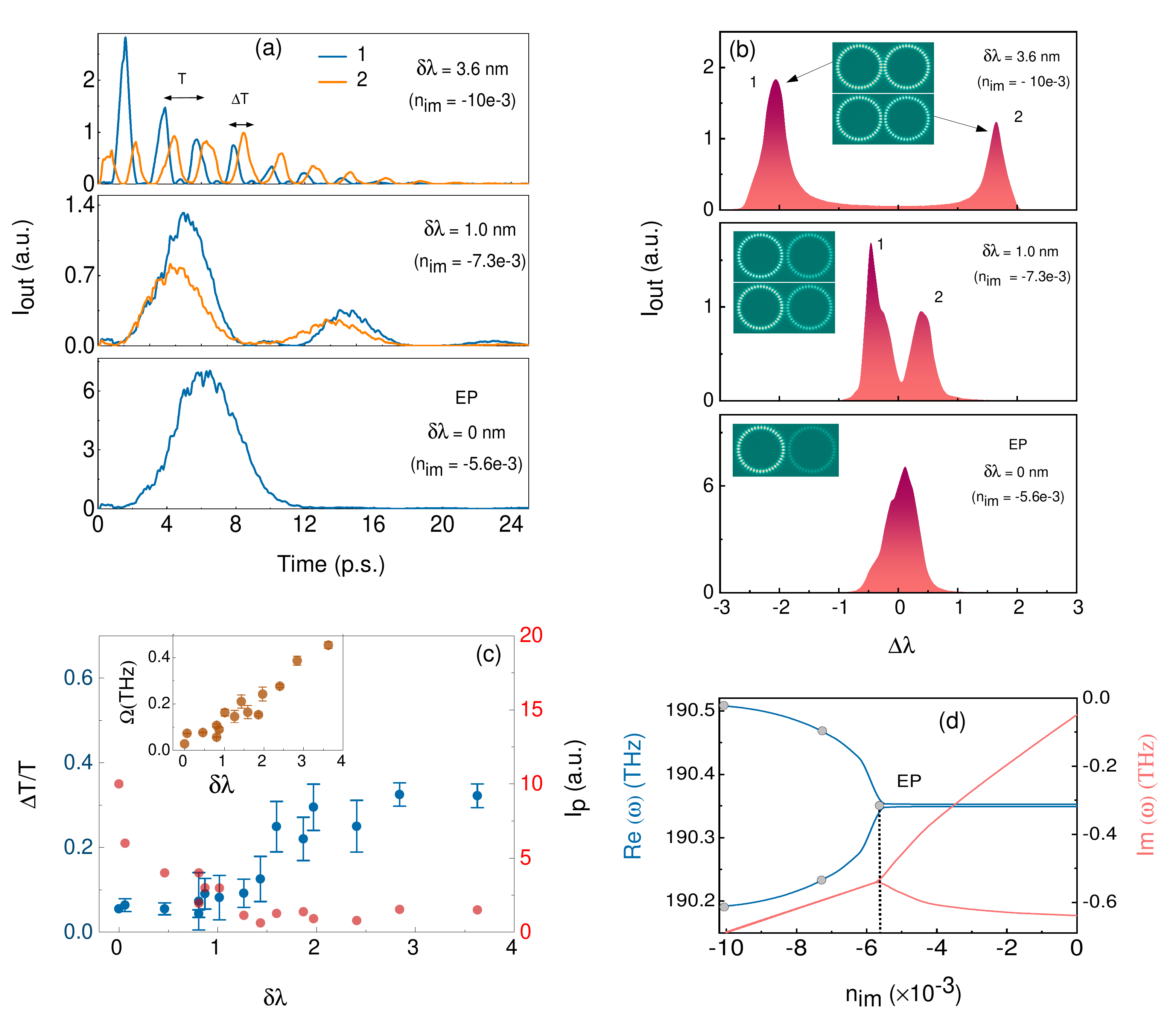}
\caption{\textbf{Quantum echo diagnosis of an exceptional point in a simulated, deterministic, model system}: (a) Temporal evolution of emission from two coupled microring resonators, as computed from finite-difference time-domain computations. Top panel: Situation for  spectral splitting of $\delta\lambda  = 3.6~$nm. Blue curve shows the echo in the emission at one wavelength, showing a period of $T$ ($=\frac{1}{\Omega}$, where $\Omega$ is the echo frequency). The echo in the emission at the other wavelength (orange curve) is out of phase, quantified by $\Delta T/T$. Middle panel: Dynamics for $\delta\lambda  = 1~$nm, where $T$ is increased, while the $\Delta T/T$ is reduced. Bottom panel: Behavior at the exceptional point (EP), i.e, $\delta\lambda  = 0~$nm, where there is no echo or oscillation. A single temporal profile is seen because of the eigenvalue coalescence. (b) Spectro-spatial observations corresponding to (a). Insets show intensity distribution in the coupled system averaged over the entire echo. The energy resides in both the resonators at large splittings, while it exists exclusively in the amplifying resonator at the EP. (c) Phase difference ($\Delta T/T$) in the two echoes (blue markers, left Y axis) reduces systematically with $\delta\lambda$, and is zero at the EP. Peak emission intensity (red markers, right Y-axis) $I_p$ diverges at the EP. Inset shows $\Omega$ with $\delta\lambda$. (d) Correspondence between the quantum echo behavior of (a) and PT-symmetry breaking and EP-formation. Circle markers mark the situations in (a) and (b).}
\label{fig:simring}
\end{figure}

We first illustrate the  quantum echo in a known EP system, namely, a coupled microring cavity. We simulate identical resonators with a diameter of $d = 2.0~\mu$m, thickness $t = 0.2~\mu$m and a separation $s = 0.4~\mu m$. The system was simulated using a finite-difference time-domain scheme (Lumerical) in the time domain, followed by full-wave finite-element computation (COMSOL) in the frequency domain. Optical gain or loss is introduced in the complex refractive index $n_{im}$ (see Methods for details). The $n_{im}$ in the second cavity is employed as the tunable parameter to approach an exceptional point, and is set to -0.01 (amplification) in the first cavity.  Fig\ref{fig:simring}(a) depicts the temporal behavior of the system, while (b) describes the corresponding eigenvalue behavior, along with the intensity distribution in the coupled resonators. The top panels of (a) and (b) depict the situation for identical gain, $n_{im} = -0.01$, in the two cavities. The resulting strong coupling realizes large splitting corresponding to a wavelength splitting $\delta\lambda$ of 3.6~nm. The emission clearly shows QE oscillations (top panel of Fig\ref{fig:simring}(a)) in the temporal profile. The blue and orange curves represent emission from the two cavities. The high QE frequency ($\Omega$) allows for multiple oscillations during the emission cycle, with out-of-phase oscillations from the two cavities. The corresponding split eigenvalues are shown in the top panel of (b), along with the intensity distribution (inset) in the cavities. The intensity is shown here instead of fields because the former is an experimental observable. Both the supermodes corresponding to the two eigenvalues show identical intensity, although the fields depict the bonding and antibonding character. See Supplementary information for the field distributions. With reducing gain, the splitting reduces proportionately. For example, the middle panel of (a) shows the situation for $n_{im} = -0.0073$ in cavity 2, where the $\delta\lambda = 1~$nm. The QE oscillation frequency and the phase difference ($\Delta T/ T$) diminishes. The intensity distribution for the two supermodes is skewed in favour of the cavity 1, as seen in (b).  Ultimately, at $n_{im} =  -0.0056$ in cavity 2, the system hits an exceptional point. The temporal emission coalesces into a single profile seen in the bottom panel of (a). The two eigenvalues coalesce and a single spectral peak is observed. The entire energy is accumulated in the cavity 1, which is indicative of eigenvector coalescence seen in the field distributions. See Supplementary Information for the fields. Fig\ref{fig:simring}(c) describes the systematic variation of the phase difference between two echoes and the output intensity with $\delta\lambda$.  A monotonic decrease in $\Delta T/T$ (blue markers, left Y-axis) with $\delta\lambda$ is seen. The peak output intensity $I_p$ increases upon approaching the EP, and maximises at the EP, indicative of enhanced functionality of the system operating at the exceptional point. The inset shows that $\Omega$ reduces monotonically with $\delta\lambda$. The correspondence of the quantum echo with the PT-symmetry breaking was confirmed by calculating the complex eigenvalues of the system. These  were computed by using a finite-element eigensolver computation, wherein the exact same structure was simulated. The gain/loss parameter $n_{im}$ was same as in the FDTD. The real (blue curve) and imaginary (red curve) eigenvalues are shown in subplot (d). The circle markers on the blue curve are the situations discussed in (a) and (b). A clear correspondence between the temporal behavior and the theoretically-expected PT-symmetry breaking and EP-formation is seen. 

\begin{figure}[h!]
\centering
\includegraphics[width=0.9\textwidth]{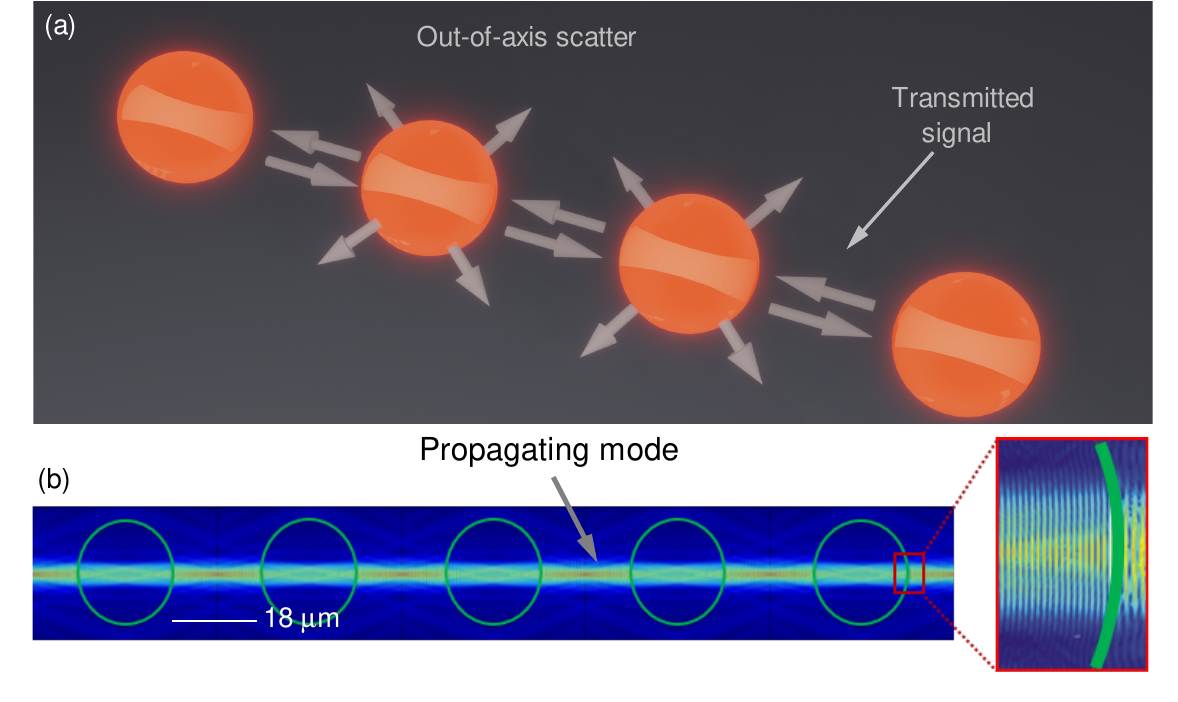}
\caption{Schematic of linear arrays of spherical microcavities: (a) white arrows show emission and scattered directions: longitudinal emission (transmitted signal) and out-of-axis emission. The sizes of the cavities are about 18~$\mu m$. (b) A Finite-Element-Method (FEM) simulation showing the longitudinal mode.}
\label{fig:1Dapprox}
\end{figure}

\begin{figure}[h!]
\centering
\includegraphics[width=1\textwidth]{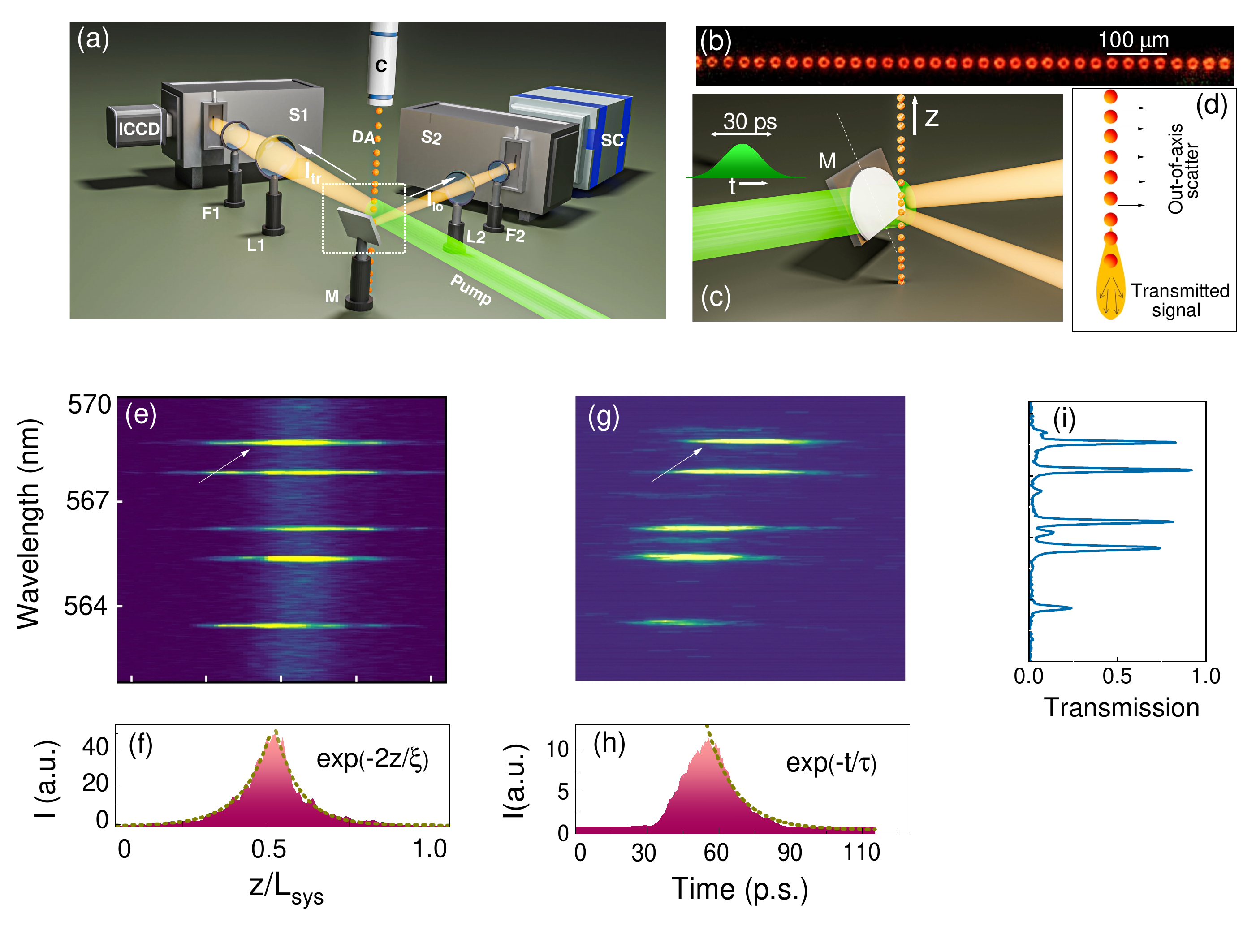}
\caption{\textbf{Experimental setup and characterization of the Anderson localized lasing system:} (a) Complete schematic of the setup, see main text for details; (b) CCD image of the microresonator array; (c) zoomed view of the rectangular region in (a): the microresonator array is pumped by a mode-locked Nd:YAG laser (green beam, $\lambda = 532~nm$, pulse width = $30~ps$, repetition rate = $10$~Hz). The emission in the longitudinal direction (transmitted light) is used for the temporal measurement $I_{tmp}$, and the transverse out-of-axis scatter  gives the spatial profile $I_{sp}$ simultaneously. The schematic of emission directions is shown in (d)). The transverse scatter is imaged onto the input slit of a spectrometer S1 coupled with ICCD using a 4f imaging setup [L2: imaging lens  (10 cm focal length); F2: Notch filter (to remove the pump beam)]. The longitudinal emission is measured using a mirror M located at an angle of $45^{0}$ w.r.t the array as shown in the schematic. This emission is directed towards the input slit of a second identical spectrometer S2 coupled to a streak camera (Optronis SC-10, temporal resolution = $\sim3$~ps)  through a focusing lens L3 and a notch filter F2.  (e, f) represent the spatial data, where (e) shows the spectrographic image of the multiple lasing modes, and (f) shows a typical extracted spatial profile of one mode, yielding localization length; (g) shows the simultaneous streak image of the same modes with (h) showing a typical extracted temporal profile. (i) Lasing spectrum obtained by integrating the streak image over time axis, or integrating the spectrographic image over spatial axis. }
\label{fig:expsetup}
\end{figure}

The structures we employ for these experiments, schematized in Fig~\ref{fig:1Dapprox}(a), are linear arrays of microcavities that couple to form photonic bands and bandgaps. The microcavities are spherical in shape with diameters of about $18~\mu$m. The large diameters of the spheres and the linear configuration of the array induces coupling of longitudinal Fabry-Perot modes of the microcavities.  Figure~\ref{fig:1Dapprox}(b) illustrates a finite-element computation of an eigenmode of the periodic array, clearly depicting the longitudinal mode with quasi-planar wavefronts, akin to a multilayer system. Loss in this structure is induced by out-of-axis scattering of the mode by the interfaces. In order to introduce gain, we choose the  material appropriately to make the microcavities, as described in the Methods.  Disorder is introduced by deliberately modifying the axial dimensions and spatial separations of the microcavities. In our earlier works, we have shown that the periodic arrays sustain Bloch modes\cite{joshi2022anomalous}, while the disordered arrays realize Anderson localized modes\cite{joshi2020reduction, joshi2019effect, kumar2017temporal}. As the fabrication process of the microcavity arrays is dynamic in nature, we can monitor a huge number of configurations, as discussed next.

Figure~\ref{fig:expsetup} illustrates the creation of the microcavity array, and its simultaneous spectro-spatio-temporal characterization. The generation technique, as described in details in the Methods section, creates the microcavities from a Rhodamine-in-alcohol solution. The Rhodamine molecules act as sources distributed throughout the length of the array, whilst also providing gain to the mode. [CCD image shown in (b)]. Optical excitation is provided by pulses of an Nd:YAG laser fired at 1 Hz, illustrated by the green beam. The emission transverse to the array comprises the out-of-axis scatter that is imaged onto a identical spectrometer (S1) coupled to an imaging CCD. The longitudinal emission of the array, which carries the temporal signature of the emission in Anderson localized modes, is directed to an identical spectrometer (S2) coupled to a Streak Camera SC (Optronis SC-10).  Since the microcavities are constantly created and introduced into the pump laser focus, every excitation laser pulse sees a different disorder configuration of the array. This feature enables us to generate the vast ensemble of configurations required in this study. Importantly, the Streak Camera and Spectrometer CCD are synchronized with the pump laser to register the spectral, spatial and temporal data of every configuration simultaneously. Panels (c) and (d) illustrate the characterization scheme. Panel (e) depicts the spectrographic image from S1 at a particular pump pulse. It shows the multiple emission modes from the corresponding configuration illuminated by the laser pulse. The horizontal axis represents the spatial axis, while the  vertical axis represents the wavelength. Subplot (f) shows the extracted spatial profile of a representative mode labeled by the white arrow in the image (e). This is an Anderson-localized mode, whose localization length $\xi$ can be extracted from the decay in the wings. Panel (g) shows the Streak image of the emission of the same configuration. Since this is directed through a spectrometer, the  vertical axis  corresponds to the wavelength, while the  horizontal axis corresponds the time axis. A comparison between (g) and (e) shows that the same modes are captured in both devices, endorsing the simultaneous diagnostic capability of the setup. Subplot (h) is the extracted temporal profile of the same mode as in (f). Subplot (i) shows the emission spectrum obtained by integrating over the time-axis of (g). The same can be obtained by integrating the space-axis of (e). 

We used this diagnostic system to characterize 3,000 different configurations of disorder, providing a massive ensemble of about 15,000 Anderson-localized lasing modes. The limit on the configuration number is set by the volume of the liquid present in the generator, as mentioned in the Methods. Although all modes were captured at the same pump energy, inherent pump fluctuations motivated us to further process the ensemble to choose modes in a very narrow range of pump energy, to within $\sim1~\mu$J at the average pump energy of $\sim87~\mu J$. The filtered ensemble still had over 5,000 modes. Among all the captured modes, pairs of coupled modes were identified through the temporal profiles, since the coupling resulted in a quantum echo. This clearly stood apart against the smooth profiles of individual uncoupled modes, such as the one shown in Figure \ref{fig:expsetup}(h). The coupling was further corroborated by the splitting in the spectral profile. Various degrees of splitting allowed us to identify the approach to a second-order exceptional point, and eventually the EP itself. 

\begin{figure}[h!]
\centering
\includegraphics[width=1\textwidth]{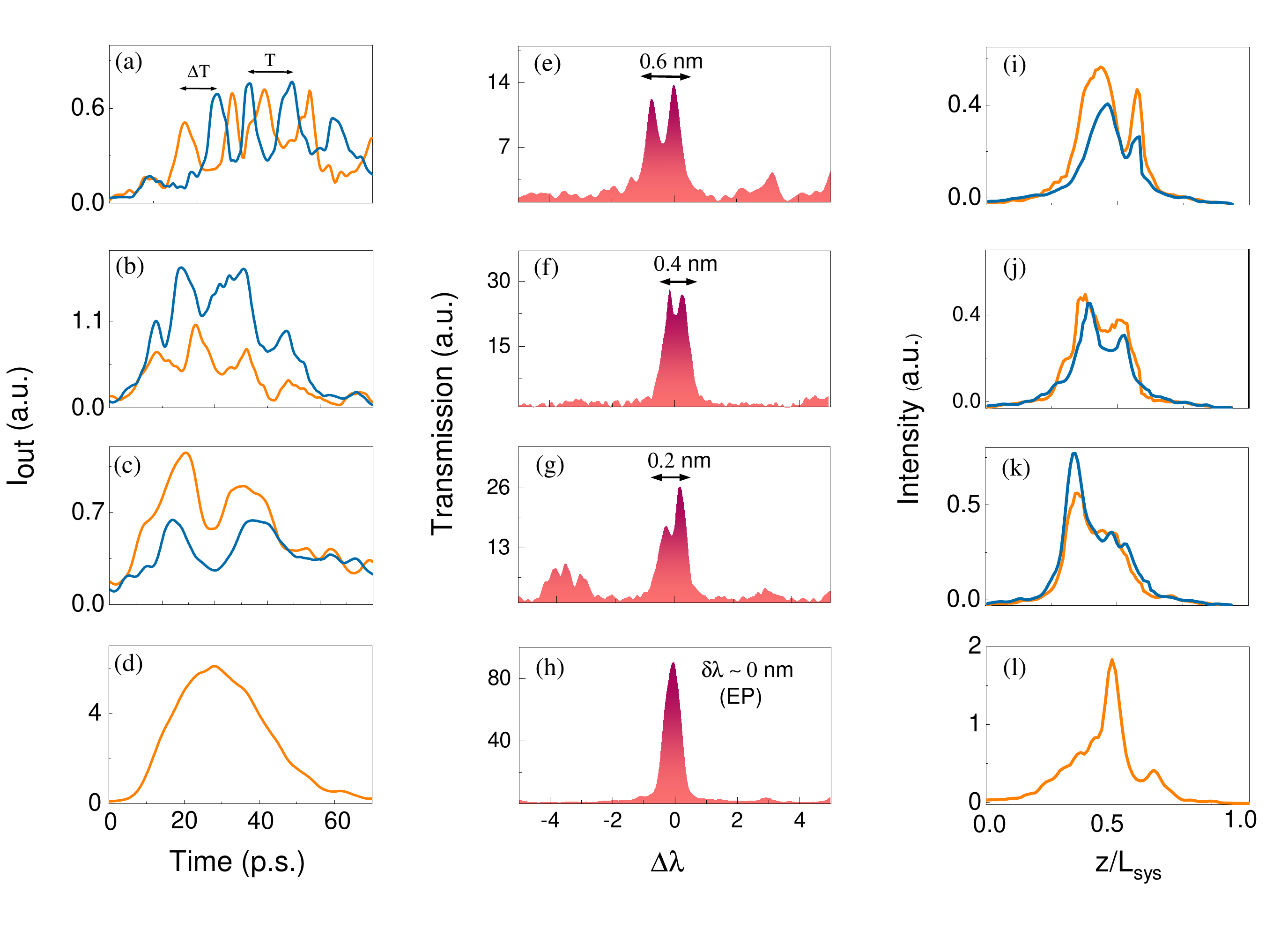}
\caption{\textbf{Experimentally measured temporal, spectral, and spatial characteristics of Anderson localized lasing modes around and at an exceptional point (EP):} Subplots (a-d) show the measured quantum echo from lasing emission, with (d) being the EP. Unlike Figure ~3(h), the temporal profiles in (a) exhibit a quantum echo that decisively certifies that the two modes are coupled. Reducing echo frequency with decreasing phase difference between the echoes at the two wavelengths is clearly observed. Subplots (e-h) depict the corresponding eigenvalue spectra from the same configurations as (a-d) with mentioned splittings. The echo behavior is clearly consistent with the splitting. (i-l)  represent corresponding spatial intensity distribution of the coupled localized states. Each spatial mode exhibits two peaks, and signatures of shape exchange. Energy is distributed in both Anderson cavities at large splittings, but resides mostly in one cavity at $\delta\lambda = 0.2$~nm. A single profile exists at the EP in (l). The intensity behavior certifies coalescence of eigenvectors towards and at the EP. Subplots (d), (h) and (l) represent Anderson-localization lasing over an exceptional point, further confirmed in Figure 5.}
\label{fig:sptemp}
\end{figure}

Figure~\ref{fig:sptemp} illustrates a representative set of measurements in the Anderson-localizing system showing the approach to an EP. The first column of panels shows the measured quantum echo, the middle column shows the corresponding spectral measurement, while the third column illustrates the corresponding spatial intensity distributions. Each row of three panels show simultaneous diagnostics from the same disorder configuration, while different rows represent data from different configurations. The subplot (a) shows a temporal profile in which a high frequency quantum echo is observed. The two profiles (orange and blue) arising from two coupled modes are out of phase by $\Delta T/T$ as indicated in the plot. Subplot(e) shows the spectral splitting in the emission from the two coupled modes in this configuration. Quite symmetric spectral peaks are observed. The marked $\delta\lambda = 0.6$~nm was measured using a peakfinder algorithm. Subplot (i) shows the spatial intensity of the two modes extracted from the spectrographic data as described earlier. Strong spatial overlap with signatures of separated peaks are seen in the profiles. The coupled Anderson localized modes are spread over almost the system length. Two distinct peaks in both the profiles are seen, indicative of the two resonant cavities that underwent coupling. These profiles strongly emphasize the fact that the two coupled cavities are not physically separated, unlike conventional photonic coupled-cavities (Figure 1). In such a disordered system, the gain/loss values in the two cavities are self-determined by the random configuration that decides the relevant parameters such as quality factors, pump distributions etc. The lower rows show gradually decreasing quantum echo frequency, which corresponds to the reduction in splitting seen in the middle panels. At the same time, the intensity distributions morphologically coalesce as the splitting diminishes. Notably, subplot (k) shows a strong peak accompanied by a shoulder, that can be interpreted as the intensity preferentially residing in one cavity. These configurations represent the approach to a second-order exception point. Ultimately, the bottom row (d, h, l) represents lasing {\it at} the exceptional point, showing only one echoless temporal profile, one spectral peak and one intensity profile. Subplots (a) through (d) illustrate the vanishing of the echo as the EP is hit. The conclusive inference that (d) corresponds to a second-order EP is provided in the next figure where we present statistical analysis of the data. 

\begin{figure}[h!]
\centering
\includegraphics[width=1\textwidth]{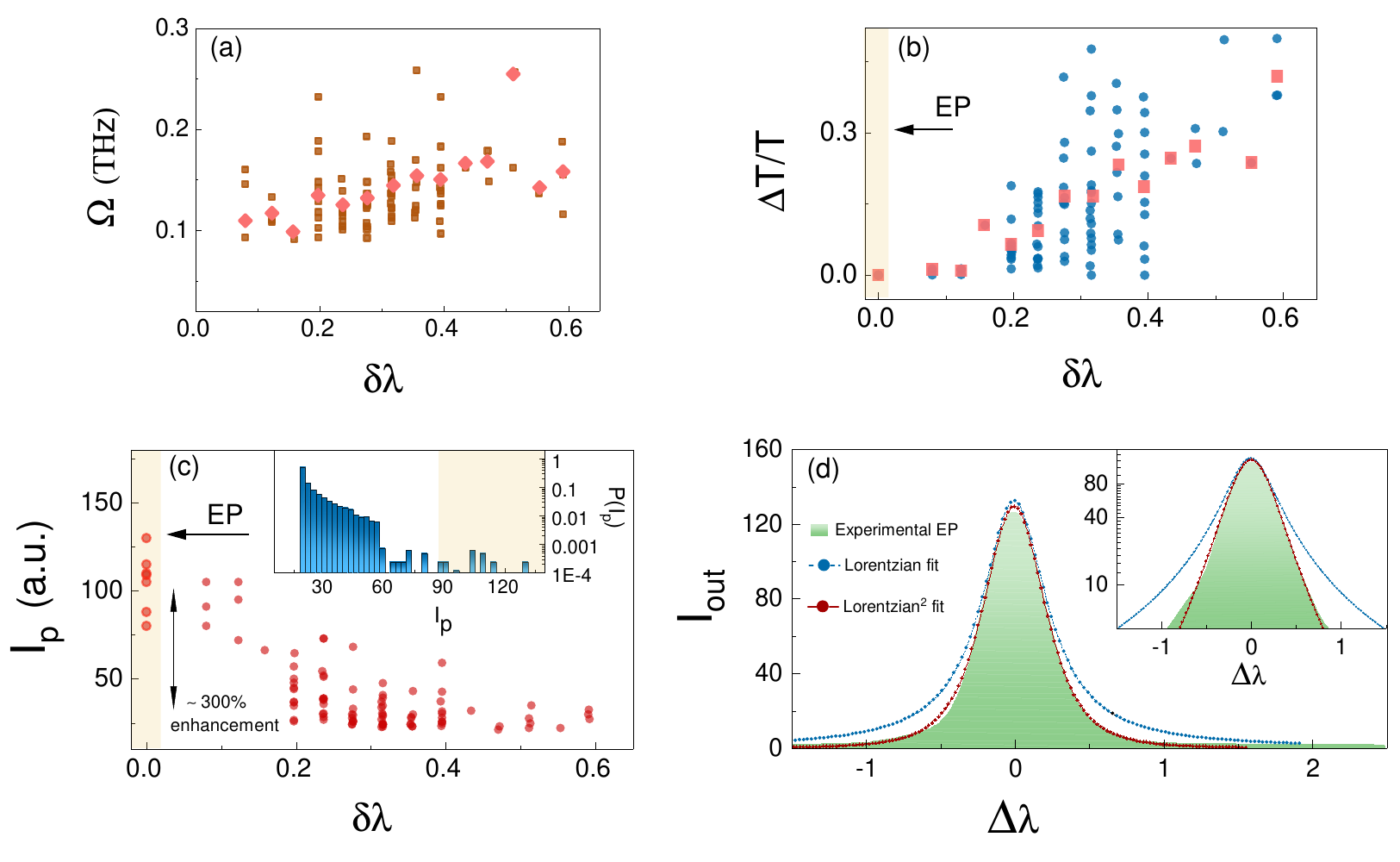}
\caption{\textbf{Confirmation of the EP in a statistical ensemble:} (a) Scatter points for the QE frequency $\Omega$ show an overall reduction with decreasing $\delta\lambda$. Pink diamonds indicate $\langle \Omega \rangle$ at a particular $\delta\lambda$. (b) Measured $\Delta T/T$ and (c) peak emission intensity ($I_p$) as a function of $\delta\lambda$, measured over $\sim100$ modes. Each datapoint in (b) is an average over a few oscillations in the temporal profile. Pink squares indicate $\langle \Delta T/T \rangle$  at a particular $\delta\lambda$. The scatter clearly shows an approach to  $\Delta T/T \rightarrow 0$ as $\delta\lambda \rightarrow 0$. The yellow band around $\delta\lambda = 0$ signifies the EP region. All datapoints here are placed at $\delta\lambda = 0$ as the $\delta\lambda$ is beyond experimental spectral resolution. Although $\Delta T/T$ does not exist here, a suggestive datapoint is marked at $\Delta T/T = 0$ in logical follow-up to the neighbouring datapoints. (c) shows increasing peak intensity $I_p$ towards the EP region, with significant jump of over $300\%$ in the $I_{p}$ in the EP region. Inset shows $P(I_p)$ with emphasis shading on outlier points of high intensity. (d) Spectral line shape at the EP (filled green plot). Blue and red plots are Lorentzian and squared-Lorentzian fits respectively. The profile possesses a clear squared-Lorentzian character, which is the classic signature of an EP. Inset shows the same in logarithmic Y axis.}
\label{fig:lorentz}
\end{figure}

Fig~\ref{fig:lorentz}(a) illustrates the frequency of the quantum echo for various splittings. The mean frequency $\langle \Omega \rangle$ at each $\delta\lambda$ is marked as pink diamonds. The trend in $\langle \Omega \rangle$ is determined by the distribution of the scatter points which are constrained to a lower limit imposed by the total pulse duration ($\sim 50$~ps). (b) shows the phase difference between the echograms of the coupled modes as a function of $\delta\lambda$ over several configurations.  The scatter in the data reflects the probabilistic variation in the phase difference for a given splitting, with a large scatter existing in the range $0.2$~nm$<\delta\lambda<0.4$~nm, wherein a large number of coupled modes were observed. Such statistical variations are expected in non-deterministic systems due to configurational variations. The $\langle \Delta T/T \rangle$ is marked as pink squares. As $\delta\lambda \rightarrow 0$~nm, the $\langle \Delta T/T \rangle \rightarrow 0$. The $\langle\delta T/T\rangle$ shows a steady rise with $\delta\lambda$. The yellow band at $\delta\lambda = 0$ signifies the EP range. Clearly, $\delta T/T$ is undefined here since there is only one mode and one temporal profile. Nonetheless, a datapoint is intentionally marked to signify the completeness of the trend. Figure \ref{fig:lorentz}(c) shows the peak intensity of the modes. The intensity rises as modes begin to coalesce spectrally, in agreement with the typical behavior in an EP-approach. Inside the EP range (yellow band at $\delta\lambda = 0$), six modes were identified that lased at a very high intensity. Upto 300\% enhancement in lasing intensity was observed at these points. Such high intensity is an indicator of enhanced functionality, a preliminary identifier of a possible EP. The inset shows the intensity distribution over the entire ensemble of modes, emphasizing the outlier nature of the high-intensity points. Finally, subplot (d) confirms the exceptional point character of the modes using an analytical fit for the lineshape of one of the high-intensity peaks. It is known that LDOS is modified at the exceptional point such that the resonant lineshape approaches to the $n^{th}$-power of Lorentzian for an $n^{th}$-order EP. Consistent with this, the red solid line in (d) depicts the square-Lorentzian profile $\propto \frac{\gamma^2}{4\pi[(\Delta\omega^2 + \gamma^2)]^2}$ that perfectly fits the experimental peak. In comparison, the blue line is the conventional Lorentzian fit which visibly deviates from the observed data. The inset shows the same on a logarithmic Y-axis, emphasizing the fit by the square-Lorentzian. The modified LDOS at the EP leads to a maximal peak enhancement of 4 in passive systems, and can be even more in the presence of gain. Although our system is random and every configuration has cavities of varying quality-factors, the average behavior of $I_p$ in subplot (c) does exhibit instances of peak enhancement $>4$ compared to the coupled modes. The Supplementary material elucidates further data regarding fits on other exceptional point modes. Thus, these spectral lineshape diagnostics confirm that the lasing modes shown here occurred over exceptional points in non-Hermitian Anderson-localizing systems. 

In summary, we have identified exceptional points in one-dimensional Anderson localizing systems, and have extracted lasing over the corresponding modes. The methodology adopted here significantly deviates from the tuning-parameter mechanism that is conventionally utilized in photonics. Our technique preserves the statistical variations in disorder that are sacrosanct in the domain of mesoscopic optics. We sample thousands of disorder configurations, and identify the quantum echo which is set up whenever coupled Anderson-localized modes are realized within the structures. The echo provides unambiguous signatures of coupling, and sets the coupled modes apart from random neighbouring modes in the spectrum. Our method realizes a large multitude of disorder configurations and simultaneously measures spatial, spectral and temporal dynamics therein. The presence of a quantum echo in the temporal signal allows us to track configurations with spectral splitting. We find that reduced splitting is accompanied by enhanced output intensity which allows us to identify individual modes that provide exceedingly large outcoupled intensity. Spectral shape analysis of these intense modes reveal the characteristic square-Lorentzian shapes originating from a second-order exceptional point degeneracy, as against Lorentzian peaks otherwise. The excess intensity in these modes is testimonial to the enhanced functionality afforded by the exceptional points.  

\section*{Methods}

\subsection*{Sample preparation and system characterization}
The system comprises an array of amplifying micro-droplets made of Rhodamine-6G (with a concentration of 1.0~mM) dissolved in a mixture of methanol and ethylene glycol with an equal proportion(50:50$\%$). Ethylene glycol(EG) is added to obtain good stability and high index contrast (refractive index of the droplet, $n_{d}$ = 1.38) due to its high viscosity and refractive index value. Rh6G has a broadband gain profile in the range of 540$-$610~nm.

The array is generated using a vibrating orifice aerosol generator (VOAG). The liquid from a chamber (1000~ml liquid capacity) was pressurized using a dry-nitrogen cylinder and passed through VOAG. The VOAG has a circular opening with a diameter of $10~\mu m$, which generates an inhomogeneous and unstable cylindrical jet of droplets. In order to make stable and spherical-shaped droplets,
a piezoelectric gate is attached at the circular opening of the VOAG. The gate is then perturbed by applying a periodic electronic signal. Depending upon the amplitude and frequency of the signal, the size and spacing of the droplets can be varied in a controlled way which allows us to generate an ensemble of monodisperse droplets which acts as a periodic array. Controlled deviations in the electronic signal perturb the microdroplets and form a disordered array. For this work, we generate a disordered array of droplets. We made a large ensemble of disordered arrays with a mean diameter of 18.6~$\mu$m. The standard deviation in the diameters is 340.0~$nm$ which determines the strength of disorder in the array.

\subsection*{Coupled mode analysis of the binary system:}

1. Eigenvalue solver using a finite element method (FEM: COMSOL Multiphysics) was used to calculate the complex eigenfrequencies of the coupled ring resonator. Simulations were performed in 2D. The structural parameters of the coupled cavities are: Diameter of the ring $D = 2.0~\mu m$, thickness $t = 0.2~\mu m$ and separation $s = 0.4~\mu m$. The cavity resonance frequency for these parameters is $\sim$190.35~THz. The cavities' gain or loss are provided in terms
of their complex refractive indices, $n = n_{r}+ in_{im}$. The real part of the refractive index ($n_{r}$) for both cavities is set at 3.3. The separation between the cavities is also fixed at $s$, thus fixing the coupling strength ($\kappa$) between the cavities. We provide a fixed amount of gain ($n_{im} = -0.01$) in the $1^{st}$ cavity. The simulation was then performed varying $n_{im}$ of the second cavity from -0.01 to 0. During the tuning, exceptional points were identified to occur at $n_{im}\sim-0.0056$. 

2. Temporal analyses were done using Finite-difference time-domain method (Lumerical FDTD). The structural and material parameters are same as mentioned in part 1 above. In this method, we use a Lorentz gain model with a Lorentz permittivity and resonance frequency close to the cavity frequency of the ring resonator. The permittivity of the material for this method is given by the relation:
\begin{equation}
\epsilon(\omega) = \epsilon + \epsilon_{l}\omega_{0}^{2}/(\omega_{0}^{2}-2i\delta_{0}\omega-\omega^{2})
\end{equation}

The gain center frequency is set at $\omega_{0}$ (Lorentz Resonance) while the width is set as $\delta_{0}$. The strength of the gain (gain amplitude) is given by the permittivity $\epsilon_{l}$, while the sign determines whether the material has loss or gain. 

The gain level ($n_{l} = \sqrt{\epsilon_{l}}$) in the $1^{st}$ cavity is kept fixed at -0.01. It was confirmed that this value of gain did not lead to any divergence due to an exponential growth of the field. The gain level in the second cavity is gradually reduced from this value and replaced with loss. Both cavities are uniformly excited using a set of identical dipole sources at the cavity resonance frequency. The time signal of the field and intensity are measured by placing field time monitors inside the cavities. Spectra are calculated using the standard Fourier transform method. A full apodization method is used to measure the frequency-resolved temporal evolution of the cavity modes. This technique applies a window function to the fields $E(t)$ before its Fourier transform. By using this method, it is possible to calculate $E(\omega)$ from a portion of the time signal.

\bmhead{Data availability}
All data relating to the paper are available from the corresponding author upon reasonable request.

\bmhead{Code availability}
All the relevant computing codes used in this study are available from the
corresponding author upon reasonable request.

\bmhead{Supplementary information}
A Supplementary Information document has been attached with the manuscript.

\bmhead{Author contributions}
S.~M. conceived the project. S.~M. and K.~J. designed the experimental setup.  K.~J. performed the experiment, numerical simulations and analyzed the data.  K.~J. and S.~M. wrote the manuscript. S.~M. supervised the whole project.

\bmhead{Competing interests}
The authors declare no competing interests.

\bmhead{Acknowledgments}
SM would like to acknowledge the Swarnajayanti Fellowship from the Department of Science and Technology, Government of India. We acknowledge funding from the Department of Atomic Energy, Government of India (12-R $\&$ D-TFR-5.02-0200).


\begin{thebibliography}{65}

\ifx \bisbn   \undefined \def \bisbn  #1{ISBN #1}\fi
\ifx \binits  \undefined \def \binits#1{#1}\fi
\ifx \bauthor  \undefined \def \bauthor#1{#1}\fi
\ifx \batitle  \undefined \def \batitle#1{#1}\fi
\ifx \bjtitle  \undefined \def \bjtitle#1{#1}\fi
\ifx \bvolume  \undefined \def \bvolume#1{\textbf{#1}}\fi
\ifx \byear  \undefined \def \byear#1{#1}\fi
\ifx \bissue  \undefined \def \bissue#1{#1}\fi
\ifx \bfpage  \undefined \def \bfpage#1{#1}\fi
\ifx \blpage  \undefined \def \blpage #1{#1}\fi
\ifx \burl  \undefined \def \burl#1{\textsf{#1}}\fi
\ifx \doiurl  \undefined \def \doiurl#1{\url{https://doi.org/#1}}\fi
\ifx \betal  \undefined \def \betal{\textit{et al.}}\fi
\ifx \binstitute  \undefined \def \binstitute#1{#1}\fi
\ifx \binstitutionaled  \undefined \def \binstitutionaled#1{#1}\fi
\ifx \bctitle  \undefined \def \bctitle#1{#1}\fi
\ifx \beditor  \undefined \def \beditor#1{#1}\fi
\ifx \bpublisher  \undefined \def \bpublisher#1{#1}\fi
\ifx \bbtitle  \undefined \def \bbtitle#1{#1}\fi
\ifx \bedition  \undefined \def \bedition#1{#1}\fi
\ifx \bseriesno  \undefined \def \bseriesno#1{#1}\fi
\ifx \blocation  \undefined \def \blocation#1{#1}\fi
\ifx \bsertitle  \undefined \def \bsertitle#1{#1}\fi
\ifx \bsnm \undefined \def \bsnm#1{#1}\fi
\ifx \bsuffix \undefined \def \bsuffix#1{#1}\fi
\ifx \bparticle \undefined \def \bparticle#1{#1}\fi
\ifx \barticle \undefined \def \barticle#1{#1}\fi
\bibcommenthead
\ifx \bconfdate \undefined \def \bconfdate #1{#1}\fi
\ifx \botherref \undefined \def \botherref #1{#1}\fi
\ifx \url \undefined \def \url#1{\textsf{#1}}\fi
\ifx \bchapter \undefined \def \bchapter#1{#1}\fi
\ifx \bbook \undefined \def \bbook#1{#1}\fi
\ifx \bcomment \undefined \def \bcomment#1{#1}\fi
\ifx \oauthor \undefined \def \oauthor#1{#1}\fi
\ifx \citeauthoryear \undefined \def \citeauthoryear#1{#1}\fi
\ifx \endbibitem  \undefined \def \endbibitem {}\fi
\ifx \bconflocation  \undefined \def \bconflocation#1{#1}\fi
\ifx \arxivurl  \undefined \def \arxivurl#1{\textsf{#1}}\fi
\csname PreBibitemsHook\endcsname

\bibitem{bender1999pt}
\begin{barticle}
\bauthor{\bsnm{Bender}, \binits{C.M.}},
\bauthor{\bsnm{Boettcher}, \binits{S.}},
\bauthor{\bsnm{Meisinger}, \binits{P.N.}}:
\batitle{{PT}-symmetric quantum mechanics}.
\bjtitle{Journal of Mathematical Physics}
\bvolume{40}(\bissue{5}),
\bfpage{2201}--\blpage{2229}
(\byear{1999})
\end{barticle}
\endbibitem

\bibitem{bender1998real}
\begin{barticle}
\bauthor{\bsnm{Bender}, \binits{C.M.}},
\bauthor{\bsnm{Boettcher}, \binits{S.}}:
\batitle{Real spectra in non-{H}ermitian {H}amiltonians having p t symmetry}.
\bjtitle{Physical Review Letters}
\bvolume{80}(\bissue{24}),
\bfpage{5243}
(\byear{1998})
\end{barticle}
\endbibitem

\bibitem{el2018non}
\begin{barticle}
\bauthor{\bsnm{El-Ganainy}, \binits{R.}},
\bauthor{\bsnm{Makris}, \binits{K.G.}},
\bauthor{\bsnm{Khajavikhan}, \binits{M.}},
\bauthor{\bsnm{Musslimani}, \binits{Z.H.}},
\bauthor{\bsnm{Rotter}, \binits{S.}},
\bauthor{\bsnm{Christodoulides}, \binits{D.N.}}:
\batitle{Non-{H}ermitian physics and {PT} symmetry}.
\bjtitle{Nature Physics}
\bvolume{14}(\bissue{1}),
\bfpage{11}--\blpage{19}
(\byear{2018})
\end{barticle}
\endbibitem

\bibitem{ruter2010observation}
\begin{barticle}
\bauthor{\bsnm{R{\"u}ter}, \binits{C.E.}},
\bauthor{\bsnm{Makris}, \binits{K.G.}},
\bauthor{\bsnm{El-Ganainy}, \binits{R.}},
\bauthor{\bsnm{Christodoulides}, \binits{D.N.}},
\bauthor{\bsnm{Segev}, \binits{M.}},
\bauthor{\bsnm{Kip}, \binits{D.}}:
\batitle{Observation of parity--time symmetry in optics}.
\bjtitle{Nature Physics}
\bvolume{6}(\bissue{3}),
\bfpage{192}--\blpage{195}
(\byear{2010})
\end{barticle}
\endbibitem

\bibitem{regensburger2012parity}
\begin{barticle}
\bauthor{\bsnm{Regensburger}, \binits{A.}},
\bauthor{\bsnm{Bersch}, \binits{C.}},
\bauthor{\bsnm{Miri}, \binits{M.-A.}},
\bauthor{\bsnm{Onishchukov}, \binits{G.}},
\bauthor{\bsnm{Christodoulides}, \binits{D.N.}},
\bauthor{\bsnm{Peschel}, \binits{U.}}:
\batitle{Parity--time synthetic photonic lattices}.
\bjtitle{Nature}
\bvolume{488}(\bissue{7410}),
\bfpage{167}--\blpage{171}
(\byear{2012})
\end{barticle}
\endbibitem

\bibitem{jing2014pt}
\begin{barticle}
\bauthor{\bsnm{Jing}, \binits{H.}},
\bauthor{\bsnm{{\"O}zdemir}, \binits{S.}},
\bauthor{\bsnm{L{\"u}}, \binits{X.-Y.}},
\bauthor{\bsnm{Zhang}, \binits{J.}},
\bauthor{\bsnm{Yang}, \binits{L.}},
\bauthor{\bsnm{Nori}, \binits{F.}}:
\batitle{{PT}-symmetric phonon laser}.
\bjtitle{Physical Review Letters}
\bvolume{113}(\bissue{5}),
\bfpage{053604}
(\byear{2014})
\end{barticle}
\endbibitem

\bibitem{lin2011unidirectional}
\begin{barticle}
\bauthor{\bsnm{Lin}, \binits{Z.}},
\bauthor{\bsnm{Ramezani}, \binits{H.}},
\bauthor{\bsnm{Eichelkraut}, \binits{T.}},
\bauthor{\bsnm{Kottos}, \binits{T.}},
\bauthor{\bsnm{Cao}, \binits{H.}},
\bauthor{\bsnm{Christodoulides}, \binits{D.N.}}:
\batitle{Unidirectional invisibility induced by {PT}-symmetric periodic
  structures}.
\bjtitle{Physical Review Letters}
\bvolume{106}(\bissue{21}),
\bfpage{213901}
(\byear{2011})
\end{barticle}
\endbibitem

\bibitem{zhu2014p}
\begin{barticle}
\bauthor{\bsnm{Zhu}, \binits{X.}},
\bauthor{\bsnm{Ramezani}, \binits{H.}},
\bauthor{\bsnm{Shi}, \binits{C.}},
\bauthor{\bsnm{Zhu}, \binits{J.}},
\bauthor{\bsnm{Zhang}, \binits{X.}}:
\batitle{{PT}-symmetric acoustics}.
\bjtitle{Physical Review X}
\bvolume{4}(\bissue{3}),
\bfpage{031042}
(\byear{2014})
\end{barticle}
\endbibitem

\bibitem{benisty2011implementation}
\begin{barticle}
\bauthor{\bsnm{Benisty}, \binits{H.}},
\bauthor{\bsnm{Degiron}, \binits{A.}},
\bauthor{\bsnm{Lupu}, \binits{A.}},
\bauthor{\bsnm{De~Lustrac}, \binits{A.}},
\bauthor{\bsnm{Ch{\'e}nais}, \binits{S.}},
\bauthor{\bsnm{Forget}, \binits{S.}},
\bauthor{\bsnm{Besbes}, \binits{M.}},
\bauthor{\bsnm{Barbillon}, \binits{G.}},
\bauthor{\bsnm{Bruyant}, \binits{A.}},
\bauthor{\bsnm{Blaize}, \binits{S.}}, \betal:
\batitle{Implementation of {PT} symmetric devices using plasmonics: principle
  and applications}.
\bjtitle{Optics Express}
\bvolume{19}(\bissue{19}),
\bfpage{18004}--\blpage{18019}
(\byear{2011})
\end{barticle}
\endbibitem

\bibitem{ding2015coalescence}
\begin{barticle}
\bauthor{\bsnm{Ding}, \binits{K.}},
\bauthor{\bsnm{Zhang}, \binits{Z.}},
\bauthor{\bsnm{Chan}, \binits{C.T.}}:
\batitle{Coalescence of exceptional points and phase diagrams for
  one-dimensional {PT}-symmetric photonic crystals}.
\bjtitle{Physical Review B}
\bvolume{92}(\bissue{23}),
\bfpage{235310}
(\byear{2015})
\end{barticle}
\endbibitem

\bibitem{peng2014parity}
\begin{barticle}
\bauthor{\bsnm{Peng}, \binits{B.}},
\bauthor{\bsnm{{\"O}zdemir}, \binits{{\c{S}}.K.}},
\bauthor{\bsnm{Lei}, \binits{F.}},
\bauthor{\bsnm{Monifi}, \binits{F.}},
\bauthor{\bsnm{Gianfreda}, \binits{M.}},
\bauthor{\bsnm{Long}, \binits{G.L.}},
\bauthor{\bsnm{Fan}, \binits{S.}},
\bauthor{\bsnm{Nori}, \binits{F.}},
\bauthor{\bsnm{Bender}, \binits{C.M.}},
\bauthor{\bsnm{Yang}, \binits{L.}}:
\batitle{Parity--time-symmetric whispering-gallery microcavities}.
\bjtitle{Nature Physics}
\bvolume{10}(\bissue{5}),
\bfpage{394}--\blpage{398}
(\byear{2014})
\end{barticle}
\endbibitem

\bibitem{peng2014loss}
\begin{barticle}
\bauthor{\bsnm{Peng}, \binits{B.}},
\bauthor{\bsnm{{\"O}zdemir}, \binits{{\c{S}}.}},
\bauthor{\bsnm{Rotter}, \binits{S.}},
\bauthor{\bsnm{Yilmaz}, \binits{H.}},
\bauthor{\bsnm{Liertzer}, \binits{M.}},
\bauthor{\bsnm{Monifi}, \binits{F.}},
\bauthor{\bsnm{Bender}, \binits{C.}},
\bauthor{\bsnm{Nori}, \binits{F.}},
\bauthor{\bsnm{Yang}, \binits{L.}}:
\batitle{Loss-induced suppression and revival of lasing}.
\bjtitle{Science}
\bvolume{346}(\bissue{6207}),
\bfpage{328}--\blpage{332}
(\byear{2014})
\end{barticle}
\endbibitem

\bibitem{cerjan2016exceptional}
\begin{barticle}
\bauthor{\bsnm{Cerjan}, \binits{A.}},
\bauthor{\bsnm{Raman}, \binits{A.}},
\bauthor{\bsnm{Fan}, \binits{S.}}:
\batitle{Exceptional contours and band structure design in parity-time
  symmetric photonic crystals}.
\bjtitle{Physical Review Letters}
\bvolume{116}(\bissue{20}),
\bfpage{203902}
(\byear{2016})
\end{barticle}
\endbibitem

\bibitem{feng2013experimental}
\begin{barticle}
\bauthor{\bsnm{Feng}, \binits{L.}},
\bauthor{\bsnm{Xu}, \binits{Y.-L.}},
\bauthor{\bsnm{Fegadolli}, \binits{W.S.}},
\bauthor{\bsnm{Lu}, \binits{M.-H.}},
\bauthor{\bsnm{Oliveira}, \binits{J.E.}},
\bauthor{\bsnm{Almeida}, \binits{V.R.}},
\bauthor{\bsnm{Chen}, \binits{Y.-F.}},
\bauthor{\bsnm{Scherer}, \binits{A.}}:
\batitle{Experimental demonstration of a unidirectional reflectionless
  parity-time metamaterial at optical frequencies}.
\bjtitle{Nature materials}
\bvolume{12}(\bissue{2}),
\bfpage{108}--\blpage{113}
(\byear{2013})
\end{barticle}
\endbibitem

\bibitem{jing2017high}
\begin{barticle}
\bauthor{\bsnm{Jing}, \binits{H.}},
\bauthor{\bsnm{{\"O}zdemir}, \binits{{\c{S}}.}},
\bauthor{\bsnm{L{\"u}}, \binits{H.}},
\bauthor{\bsnm{Nori}, \binits{F.}}:
\batitle{High-order exceptional points in optomechanics}.
\bjtitle{Scientific Reports}
\bvolume{7}(\bissue{1}),
\bfpage{1}--\blpage{10}
(\byear{2017})
\end{barticle}
\endbibitem

\bibitem{ding2016emergence}
\begin{barticle}
\bauthor{\bsnm{Ding}, \binits{K.}},
\bauthor{\bsnm{Ma}, \binits{G.}},
\bauthor{\bsnm{Xiao}, \binits{M.}},
\bauthor{\bsnm{Zhang}, \binits{Z.}},
\bauthor{\bsnm{Chan}, \binits{C.T.}}:
\batitle{Emergence, coalescence, and topological properties of multiple
  exceptional points and their experimental realization}.
\bjtitle{Physical Review X}
\bvolume{6}(\bissue{2}),
\bfpage{021007}
(\byear{2016})
\end{barticle}
\endbibitem

\bibitem{kang2016chiral}
\begin{barticle}
\bauthor{\bsnm{Kang}, \binits{M.}},
\bauthor{\bsnm{Chen}, \binits{J.}},
\bauthor{\bsnm{Chong}, \binits{Y.}}:
\batitle{Chiral exceptional points in metasurfaces}.
\bjtitle{Physical Review A}
\bvolume{94}(\bissue{3}),
\bfpage{033834}
(\byear{2016})
\end{barticle}
\endbibitem

\bibitem{zhen2015spawning}
\begin{barticle}
\bauthor{\bsnm{Zhen}, \binits{B.}},
\bauthor{\bsnm{Hsu}, \binits{C.W.}},
\bauthor{\bsnm{Igarashi}, \binits{Y.}},
\bauthor{\bsnm{Lu}, \binits{L.}},
\bauthor{\bsnm{Kaminer}, \binits{I.}},
\bauthor{\bsnm{Pick}, \binits{A.}},
\bauthor{\bsnm{Chua}, \binits{S.-L.}},
\bauthor{\bsnm{Joannopoulos}, \binits{J.D.}},
\bauthor{\bsnm{Solja{\v{c}}i{\'c}}, \binits{M.}}:
\batitle{Spawning rings of exceptional points out of dirac cones}.
\bjtitle{Nature}
\bvolume{525}(\bissue{7569}),
\bfpage{354}--\blpage{358}
(\byear{2015})
\end{barticle}
\endbibitem

\bibitem{heiss2012physics}
\begin{barticle}
\bauthor{\bsnm{Heiss}, \binits{W.}}:
\batitle{The physics of exceptional points}.
\bjtitle{Journal of Physics A: Mathematical and Theoretical}
\bvolume{45}(\bissue{44}),
\bfpage{444016}
(\byear{2012})
\end{barticle}
\endbibitem

\bibitem{kato1966analytic}
\begin{botherref}
\oauthor{\bsnm{Kato}, \binits{T.}}:
Analytic perturbation theory,
pp. 364--426.
Springer
(1966)
\end{botherref}
\endbibitem

\bibitem{heiss1999phases}
\begin{barticle}
\bauthor{\bsnm{Heiss}, \binits{W.}}:
\batitle{Phases of wave functions and level repulsion}.
\bjtitle{The European Physical Journal D-Atomic, Molecular, Optical and Plasma
  Physics}
\bvolume{7}(\bissue{1}),
\bfpage{1}--\blpage{4}
(\byear{1999})
\end{barticle}
\endbibitem

\bibitem{moiseyev2011non}
\begin{botherref}
\oauthor{\bsnm{Moiseyev}, \binits{N.}}:
Non-{H}ermitian quantum mechanics.
Cambridge University Press
(2011)
\end{botherref}
\endbibitem

\bibitem{el2007theory}
\begin{barticle}
\bauthor{\bsnm{El-Ganainy}, \binits{R.}},
\bauthor{\bsnm{Makris}, \binits{K.}},
\bauthor{\bsnm{Christodoulides}, \binits{D.}},
\bauthor{\bsnm{Musslimani}, \binits{Z.H.}}:
\batitle{Theory of coupled optical {PT}-symmetric structures}.
\bjtitle{Optics Letters}
\bvolume{32}(\bissue{17}),
\bfpage{2632}--\blpage{2634}
(\byear{2007})
\end{barticle}
\endbibitem

\bibitem{pick2017general}
\begin{barticle}
\bauthor{\bsnm{Pick}, \binits{A.}},
\bauthor{\bsnm{Zhen}, \binits{B.}},
\bauthor{\bsnm{Miller}, \binits{O.D.}},
\bauthor{\bsnm{Hsu}, \binits{C.W.}},
\bauthor{\bsnm{Hernandez}, \binits{F.}},
\bauthor{\bsnm{Rodriguez}, \binits{A.W.}},
\bauthor{\bsnm{Solja{\v{c}}i{\'c}}, \binits{M.}},
\bauthor{\bsnm{Johnson}, \binits{S.G.}}:
\batitle{General theory of spontaneous emission near exceptional points}.
\bjtitle{Optics Express}
\bvolume{25}(\bissue{11}),
\bfpage{12325}--\blpage{12348}
(\byear{2017})
\end{barticle}
\endbibitem

\bibitem{miri2019exceptional}
\begin{barticle}
\bauthor{\bsnm{Miri}, \binits{M.-A.}},
\bauthor{\bsnm{Al{\`u}}, \binits{A.}}:
\batitle{Exceptional points in optics and photonics}.
\bjtitle{Science}
\bvolume{363}(\bissue{6422}),
\bfpage{7709}
(\byear{2019})
\end{barticle}
\endbibitem

\bibitem{moiseyev1998quantum}
\begin{barticle}
\bauthor{\bsnm{Moiseyev}, \binits{N.}}:
\batitle{Quantum theory of resonances: calculating energies, widths and
  cross-sections by complex scaling}.
\bjtitle{Physics Reports}
\bvolume{302}(\bissue{5-6}),
\bfpage{212}--\blpage{293}
(\byear{1998})
\end{barticle}
\endbibitem

\bibitem{feng2014single}
\begin{barticle}
\bauthor{\bsnm{Feng}, \binits{L.}},
\bauthor{\bsnm{Wong}, \binits{Z.J.}},
\bauthor{\bsnm{Ma}, \binits{R.-M.}},
\bauthor{\bsnm{Wang}, \binits{Y.}},
\bauthor{\bsnm{Zhang}, \binits{X.}}:
\batitle{Single-mode laser by parity-time symmetry breaking}.
\bjtitle{Science}
\bvolume{346}(\bissue{6212}),
\bfpage{972}--\blpage{975}
(\byear{2014})
\end{barticle}
\endbibitem

\bibitem{hodaei2014parity}
\begin{barticle}
\bauthor{\bsnm{Hodaei}, \binits{H.}},
\bauthor{\bsnm{Miri}, \binits{M.-A.}},
\bauthor{\bsnm{Heinrich}, \binits{M.}},
\bauthor{\bsnm{Christodoulides}, \binits{D.N.}},
\bauthor{\bsnm{Khajavikhan}, \binits{M.}}:
\batitle{Parity-time--symmetric microring lasers}.
\bjtitle{Science}
\bvolume{346}(\bissue{6212}),
\bfpage{975}--\blpage{978}
(\byear{2014})
\end{barticle}
\endbibitem

\bibitem{takata2021observing}
\begin{barticle}
\bauthor{\bsnm{Takata}, \binits{K.}},
\bauthor{\bsnm{Nozaki}, \binits{K.}},
\bauthor{\bsnm{Kuramochi}, \binits{E.}},
\bauthor{\bsnm{Matsuo}, \binits{S.}},
\bauthor{\bsnm{Takeda}, \binits{K.}},
\bauthor{\bsnm{Fujii}, \binits{T.}},
\bauthor{\bsnm{Kita}, \binits{S.}},
\bauthor{\bsnm{Shinya}, \binits{A.}},
\bauthor{\bsnm{Notomi}, \binits{M.}}:
\batitle{Observing exceptional point degeneracy of radiation with electrically
  pumped photonic crystal coupled-nanocavity lasers}.
\bjtitle{Optica}
\bvolume{8}(\bissue{2}),
\bfpage{184}--\blpage{192}
(\byear{2021})
\end{barticle}
\endbibitem

\bibitem{hodaei2017enhanced}
\begin{barticle}
\bauthor{\bsnm{Hodaei}, \binits{H.}},
\bauthor{\bsnm{Hassan}, \binits{A.U.}},
\bauthor{\bsnm{Wittek}, \binits{S.}},
\bauthor{\bsnm{Garcia-Gracia}, \binits{H.}},
\bauthor{\bsnm{El-Ganainy}, \binits{R.}},
\bauthor{\bsnm{Christodoulides}, \binits{D.N.}},
\bauthor{\bsnm{Khajavikhan}, \binits{M.}}:
\batitle{Enhanced sensitivity at higher-order exceptional points}.
\bjtitle{Nature}
\bvolume{548}(\bissue{7666}),
\bfpage{187}--\blpage{191}
(\byear{2017})
\end{barticle}
\endbibitem

\bibitem{chen2017exceptional}
\begin{barticle}
\bauthor{\bsnm{Chen}, \binits{W.}},
\bauthor{\bsnm{Kaya~{\"O}zdemir}, \binits{{\c{S}}.}},
\bauthor{\bsnm{Zhao}, \binits{G.}},
\bauthor{\bsnm{Wiersig}, \binits{J.}},
\bauthor{\bsnm{Yang}, \binits{L.}}:
\batitle{Exceptional points enhance sensing in an optical microcavity}.
\bjtitle{Nature}
\bvolume{548}(\bissue{7666}),
\bfpage{192}--\blpage{196}
(\byear{2017})
\end{barticle}
\endbibitem

\bibitem{hokmabadi2019non}
\begin{barticle}
\bauthor{\bsnm{Hokmabadi}, \binits{M.P.}},
\bauthor{\bsnm{Schumer}, \binits{A.}},
\bauthor{\bsnm{Christodoulides}, \binits{D.N.}},
\bauthor{\bsnm{Khajavikhan}, \binits{M.}}:
\batitle{Non-{H}ermitian ring laser gyroscopes with enhanced sagnac
  sensitivity}.
\bjtitle{Nature}
\bvolume{576}(\bissue{7785}),
\bfpage{70}--\blpage{74}
(\byear{2019})
\end{barticle}
\endbibitem

\bibitem{ozdemir2019parity}
\begin{barticle}
\bauthor{\bsnm{{\"O}zdemir}, \binits{{\c{S}}.K.}},
\bauthor{\bsnm{Rotter}, \binits{S.}},
\bauthor{\bsnm{Nori}, \binits{F.}},
\bauthor{\bsnm{Yang}, \binits{L.}}:
\batitle{Parity--time symmetry and exceptional points in photonics}.
\bjtitle{Nature Materials}
\bvolume{18}(\bissue{8}),
\bfpage{783}--\blpage{798}
(\byear{2019})
\end{barticle}
\endbibitem

\bibitem{brandstetter2014reversing}
\begin{barticle}
\bauthor{\bsnm{Brandstetter}, \binits{M.}},
\bauthor{\bsnm{Liertzer}, \binits{M.}},
\bauthor{\bsnm{Deutsch}, \binits{C.}},
\bauthor{\bsnm{Klang}, \binits{P.}},
\bauthor{\bsnm{Sch{\"o}berl}, \binits{J.}},
\bauthor{\bsnm{T{\"u}reci}, \binits{H.E.}},
\bauthor{\bsnm{Strasser}, \binits{G.}},
\bauthor{\bsnm{Unterrainer}, \binits{K.}},
\bauthor{\bsnm{Rotter}, \binits{S.}}:
\batitle{Reversing the pump dependence of a laser at an exceptional point}.
\bjtitle{Nature Communications}
\bvolume{5}(\bissue{1}),
\bfpage{1}--\blpage{7}
(\byear{2014})
\end{barticle}
\endbibitem

\bibitem{vazquez2014gain}
\begin{barticle}
\bauthor{\bsnm{V{\'a}zquez-Candanedo}, \binits{O.}},
\bauthor{\bsnm{Hern{\'a}ndez-Herrej{\'o}n}, \binits{J.}},
\bauthor{\bsnm{Izrailev}, \binits{F.}},
\bauthor{\bsnm{Christodoulides}, \binits{D.}}:
\batitle{Gain-or loss-induced localization in one-dimensional {PT}-symmetric
  tight-binding models}.
\bjtitle{Physical Review A}
\bvolume{89}(\bissue{1}),
\bfpage{013832}
(\byear{2014})
\end{barticle}
\endbibitem

\bibitem{bachelard2022coalescence}
\begin{barticle}
\bauthor{\bsnm{Bachelard}, \binits{N.}},
\bauthor{\bsnm{Schumer}, \binits{A.}},
\bauthor{\bsnm{Kumar}, \binits{B.}},
\bauthor{\bsnm{Garay}, \binits{C.}},
\bauthor{\bsnm{Arlandis}, \binits{J.}},
\bauthor{\bsnm{Touzani}, \binits{R.}},
\bauthor{\bsnm{Sebbah}, \binits{P.}}:
\batitle{Coalescence of {A}nderson-localized modes at an exceptional point in
  2d random media}.
\bjtitle{Optics Express}
\bvolume{30}(\bissue{11}),
\bfpage{18098}--\blpage{18107}
(\byear{2022})
\end{barticle}
\endbibitem

\bibitem{davy2019probing}
\begin{barticle}
\bauthor{\bsnm{Davy}, \binits{M.}},
\bauthor{\bsnm{Genack}, \binits{A.Z.}}:
\batitle{Probing nonorthogonality of eigenfunctions and its impact on transport
  through open systems}.
\bjtitle{Physical Review Research}
\bvolume{1}(\bissue{3}),
\bfpage{033026}
(\byear{2019})
\end{barticle}
\endbibitem

\bibitem{huang2021wave}
\begin{barticle}
\bauthor{\bsnm{Huang}, \binits{Y.}},
\bauthor{\bsnm{Kang}, \binits{Y.}},
\bauthor{\bsnm{Genack}, \binits{A.Z.}}:
\batitle{Wave excitation and dynamics in non-{H}ermitian disordered systems}.
\bjtitle{Physical Review Research}
\bvolume{4},
\bfpage{013102}
(\byear{2022})
\end{barticle}
\endbibitem

\bibitem{weidemann2020nonhermitian}
\begin{botherref}
\oauthor{\bsnm{Weidemann}, \binits{S.}},
\oauthor{\bsnm{Kremer}, \binits{M.}},
\oauthor{\bsnm{Longhi}, \binits{S.}},
\oauthor{\bsnm{Szameit}, \binits{A.}}:
Non-{H}ermitian {A}nderson Transport
(2020)
\end{botherref}
\endbibitem

\bibitem{balasubrahmaniyam2020necklace}
\begin{barticle}
\bauthor{\bsnm{Balasubrahmaniyam}, \binits{M.}},
\bauthor{\bsnm{Mondal}, \binits{S.}},
\bauthor{\bsnm{Mujumdar}, \binits{S.}}:
\batitle{Necklace-state-mediated anomalous enhancement of transport in
  {A}nderson-localized non-{H}ermitian hybrid systems}.
\bjtitle{Physical Review Letters}
\bvolume{124}(\bissue{12}),
\bfpage{123901}
(\byear{2020})
\end{barticle}
\endbibitem

\bibitem{sahoo2022anomalous}
\begin{barticle}
\bauthor{\bsnm{Sahoo}, \binits{H.}},
\bauthor{\bsnm{Vijay}, \binits{R.}},
\bauthor{\bsnm{Mujumdar}, \binits{S.}}:
\batitle{{A}nomalous transport regime in a non-{H}ermitian {A}nderson-localized
  hybrid system}.
\bjtitle{Physical Review Research}
\bvolume{4}(\bissue{4}),
\bfpage{043081}
(\byear{2022})
\end{barticle}
\endbibitem

\bibitem{hodaei2015parity}
\begin{barticle}
\bauthor{\bsnm{Hodaei}, \binits{H.}},
\bauthor{\bsnm{Miri}, \binits{M.A.}},
\bauthor{\bsnm{Hassan}, \binits{A.U.}},
\bauthor{\bsnm{Hayenga}, \binits{W.}},
\bauthor{\bsnm{Heinrich}, \binits{M.}},
\bauthor{\bsnm{Christodoulides}, \binits{D.}},
\bauthor{\bsnm{Khajavikhan}, \binits{M.}}:
\batitle{Parity-time-symmetric coupled microring lasers operating around an
  exceptional point}.
\bjtitle{Optics Letters}
\bvolume{40}(\bissue{21}),
\bfpage{4955}--\blpage{4958}
(\byear{2015})
\end{barticle}
\endbibitem

\bibitem{kim2016direct}
\begin{barticle}
\bauthor{\bsnm{Kim}, \binits{K.-H.}},
\bauthor{\bsnm{Hwang}, \binits{M.-S.}},
\bauthor{\bsnm{Kim}, \binits{H.-R.}},
\bauthor{\bsnm{Choi}, \binits{J.-H.}},
\bauthor{\bsnm{No}, \binits{Y.-S.}},
\bauthor{\bsnm{Park}, \binits{H.-G.}}:
\batitle{Direct observation of exceptional points in coupled photonic-crystal
  lasers with asymmetric optical gains}.
\bjtitle{Nature Communications}
\bvolume{7}(\bissue{1}),
\bfpage{1}--\blpage{9}
(\byear{2016})
\end{barticle}
\endbibitem

\bibitem{chang2014parity}
\begin{barticle}
\bauthor{\bsnm{Chang}, \binits{L.}},
\bauthor{\bsnm{Jiang}, \binits{X.}},
\bauthor{\bsnm{Hua}, \binits{S.}},
\bauthor{\bsnm{Yang}, \binits{C.}},
\bauthor{\bsnm{Wen}, \binits{J.}},
\bauthor{\bsnm{Jiang}, \binits{L.}},
\bauthor{\bsnm{Li}, \binits{G.}},
\bauthor{\bsnm{Wang}, \binits{G.}},
\bauthor{\bsnm{Xiao}, \binits{M.}}:
\batitle{Parity--time symmetry and variable optical isolation in
  active--passive-coupled microresonators}.
\bjtitle{Nature Photonics}
\bvolume{8}(\bissue{7}),
\bfpage{524}--\blpage{529}
(\byear{2014})
\end{barticle}
\endbibitem

\bibitem{anderson1958absence}
\begin{barticle}
\bauthor{\bsnm{Anderson}, \binits{P.W.}}:
\batitle{Absence of diffusion in certain random lattices}.
\bjtitle{Physical Review}
\bvolume{109}(\bissue{5}),
\bfpage{1492}
(\byear{1958})
\end{barticle}
\endbibitem

\bibitem{john1987strong}
\begin{barticle}
\bauthor{\bsnm{John}, \binits{S.}}:
\batitle{Strong localization of photons in certain disordered dielectric
  superlattices}.
\bjtitle{Physical Review Letters}
\bvolume{58}(\bissue{23}),
\bfpage{2486}
(\byear{1987})
\end{barticle}
\endbibitem

\bibitem{segev2013anderson}
\begin{barticle}
\bauthor{\bsnm{Segev}, \binits{M.}},
\bauthor{\bsnm{Silberberg}, \binits{Y.}},
\bauthor{\bsnm{Christodoulides}, \binits{D.N.}}:
\batitle{{A}nderson localization of light}.
\bjtitle{Nature Photonics}
\bvolume{7}(\bissue{3}),
\bfpage{197}--\blpage{204}
(\byear{2013})
\end{barticle}
\endbibitem

\bibitem{chabanov2000statistical}
\begin{barticle}
\bauthor{\bsnm{Chabanov}, \binits{A.}},
\bauthor{\bsnm{Stoytchev}, \binits{M.}},
\bauthor{\bsnm{Genack}, \binits{A.}}:
\batitle{Statistical signatures of photon localization}.
\bjtitle{Nature}
\bvolume{404}(\bissue{6780}),
\bfpage{850}--\blpage{853}
(\byear{2000})
\end{barticle}
\endbibitem

\bibitem{wiersma1997localization}
\begin{barticle}
\bauthor{\bsnm{Wiersma}, \binits{D.S.}},
\bauthor{\bsnm{Bartolini}, \binits{P.}},
\bauthor{\bsnm{Lagendijk}, \binits{A.}},
\bauthor{\bsnm{Righini}, \binits{R.}}:
\batitle{Localization of light in a disordered medium}.
\bjtitle{Nature}
\bvolume{390}(\bissue{6661}),
\bfpage{671}--\blpage{673}
(\byear{1997})
\end{barticle}
\endbibitem

\bibitem{sapienza2010cavity}
\begin{barticle}
\bauthor{\bsnm{Sapienza}, \binits{L.}},
\bauthor{\bsnm{Thyrrestrup}, \binits{H.}},
\bauthor{\bsnm{Stobbe}, \binits{S.}},
\bauthor{\bsnm{Garcia}, \binits{P.D.}},
\bauthor{\bsnm{Smolka}, \binits{S.}},
\bauthor{\bsnm{Lodahl}, \binits{P.}}:
\batitle{Cavity quantum electrodynamics with {A}nderson-localized modes}.
\bjtitle{Science}
\bvolume{327}(\bissue{5971}),
\bfpage{1352}--\blpage{1355}
(\byear{2010})
\end{barticle}
\endbibitem

\bibitem{schwartz2007transport}
\begin{barticle}
\bauthor{\bsnm{Schwartz}, \binits{T.}},
\bauthor{\bsnm{Bartal}, \binits{G.}},
\bauthor{\bsnm{Fishman}, \binits{S.}},
\bauthor{\bsnm{Segev}, \binits{M.}}:
\batitle{Transport and {A}nderson localization in disordered two-dimensional
  photonic lattices}.
\bjtitle{Nature}
\bvolume{446}(\bissue{7131}),
\bfpage{52}--\blpage{55}
(\byear{2007})
\end{barticle}
\endbibitem

\bibitem{lahini2008anderson}
\begin{barticle}
\bauthor{\bsnm{Lahini}, \binits{Y.}},
\bauthor{\bsnm{Avidan}, \binits{A.}},
\bauthor{\bsnm{Pozzi}, \binits{F.}},
\bauthor{\bsnm{Sorel}, \binits{M.}},
\bauthor{\bsnm{Morandotti}, \binits{R.}},
\bauthor{\bsnm{Christodoulides}, \binits{D.N.}},
\bauthor{\bsnm{Silberberg}, \binits{Y.}}:
\batitle{{A}nderson localization and nonlinearity in one-dimensional disordered
  photonic lattices}.
\bjtitle{Physical Review Letters}
\bvolume{100}(\bissue{1}),
\bfpage{013906}
(\byear{2008})
\end{barticle}
\endbibitem

\bibitem{riboli2011anderson}
\begin{barticle}
\bauthor{\bsnm{Riboli}, \binits{F.}},
\bauthor{\bsnm{Barthelemy}, \binits{P.}},
\bauthor{\bsnm{Vignolini}, \binits{S.}},
\bauthor{\bsnm{Intonti}, \binits{F.}},
\bauthor{\bsnm{De~Rossi}, \binits{A.}},
\bauthor{\bsnm{Combrie}, \binits{S.}},
\bauthor{\bsnm{Wiersma}, \binits{D.}}:
\batitle{{A}nderson localization of near-visible light in two dimensions}.
\bjtitle{Optics Letters}
\bvolume{36}(\bissue{2}),
\bfpage{127}--\blpage{129}
(\byear{2011})
\end{barticle}
\endbibitem

\bibitem{joshi2020reduction}
\begin{barticle}
\bauthor{\bsnm{Joshi}, \binits{K.}},
\bauthor{\bsnm{Mondal}, \binits{S.}},
\bauthor{\bsnm{Kumar}, \binits{R.}},
\bauthor{\bsnm{Mujumdar}, \binits{S.}}:
\batitle{Reduction in generalized conductance with increasing gain in
  amplifying {A}nderson-localized systems}.
\bjtitle{Optics Letters}
\bvolume{45}(\bissue{8}),
\bfpage{2239}--\blpage{2242}
(\byear{2020})
\end{barticle}
\endbibitem

\bibitem{joshi2022anomalous}
\begin{barticle}
\bauthor{\bsnm{Joshi}, \binits{K.}},
\bauthor{\bsnm{Kumar}, \binits{R.}},
\bauthor{\bsnm{Mujumdar}, \binits{S.}}:
\batitle{Anomalous localization and transport behavior of amplifying
  periodic-on-average random systems at critical disorder}.
\bjtitle{Physical Review A}
\bvolume{105}(\bissue{3}),
\bfpage{033505}
(\byear{2022})
\end{barticle}
\endbibitem

\bibitem{dietz2007rabi}
\begin{barticle}
\bauthor{\bsnm{Dietz}, \binits{B.}},
\bauthor{\bsnm{Friedrich}, \binits{T.}},
\bauthor{\bsnm{Metz}, \binits{J.}},
\bauthor{\bsnm{Miski-Oglu}, \binits{M.}},
\bauthor{\bsnm{Richter}, \binits{A.}},
\bauthor{\bsnm{Sch{\"a}fer}, \binits{F.}},
\bauthor{\bsnm{Stafford}, \binits{C.}}:
\batitle{Rabi oscillations at exceptional points in microwave billiards}.
\bjtitle{Physical Review E}
\bvolume{75}(\bissue{2}),
\bfpage{027201}
(\byear{2007})
\end{barticle}
\endbibitem

\bibitem{dembowski2004first}
\begin{barticle}
\bauthor{\bsnm{Dembowski}, \binits{C.}},
\bauthor{\bsnm{Dietz}, \binits{B.}},
\bauthor{\bsnm{Friedrich}, \binits{T.}},
\bauthor{\bsnm{Gr{\"a}f}, \binits{H.-D.}},
\bauthor{\bsnm{Heine}, \binits{A.}},
\bauthor{\bsnm{Mej{\'\i}a-Monasterio}, \binits{C.}},
\bauthor{\bsnm{Miski-Oglu}, \binits{M.}},
\bauthor{\bsnm{Richter}, \binits{A.}},
\bauthor{\bsnm{Seligman}, \binits{T.}}:
\batitle{First experimental evidence for quantum echoes in scattering systems}.
\bjtitle{Physical Review Letters}
\bvolume{93}(\bissue{13}),
\bfpage{134102}
(\byear{2004})
\end{barticle}
\endbibitem

\bibitem{joshi2019effect}
\begin{barticle}
\bauthor{\bsnm{Joshi}, \binits{K.}},
\bauthor{\bsnm{Kumar}, \binits{R.}},
\bauthor{\bsnm{Balasubrahmaniyam}, \binits{M.}},
\bauthor{\bsnm{Mujumdar}, \binits{S.}}:
\batitle{Effect of critical disorder on lifetime distributions of
  {A}nderson-localized lasing modes}.
\bjtitle{Physical Review A}
\bvolume{100}(\bissue{2}),
\bfpage{023803}
(\byear{2019})
\end{barticle}
\endbibitem

\bibitem{kumar2017temporal}
\begin{barticle}
\bauthor{\bsnm{Kumar}, \binits{R.}},
\bauthor{\bsnm{Balasubrahmaniyam}, \binits{M.}},
\bauthor{\bsnm{Alee}, \binits{K.S.}},
\bauthor{\bsnm{Mujumdar}, \binits{S.}}:
\batitle{Temporal complexity in emission from {A}nderson localized lasers}.
\bjtitle{Physical Review A}
\bvolume{96}(\bissue{6}),
\bfpage{063816}
(\byear{2017})
\end{barticle}
\endbibitem

\bibitem{makris2008beam}
\begin{barticle}
\bauthor{\bsnm{Makris}, \binits{K.G.}},
\bauthor{\bsnm{El-Ganainy}, \binits{R.}},
\bauthor{\bsnm{Christodoulides}, \binits{D.}},
\bauthor{\bsnm{Musslimani}, \binits{Z.H.}}:
\batitle{Beam dynamics in {PT} symmetric optical lattices}.
\bjtitle{Physical Review Letters}
\bvolume{100}(\bissue{10}),
\bfpage{103904}
(\byear{2008})
\end{barticle}
\endbibitem

\bibitem{gao2015observation}
\begin{barticle}
\bauthor{\bsnm{Gao}, \binits{T.}},
\bauthor{\bsnm{Estrecho}, \binits{E.}},
\bauthor{\bsnm{Bliokh}, \binits{K.}},
\bauthor{\bsnm{Liew}, \binits{T.}},
\bauthor{\bsnm{Fraser}, \binits{M.}},
\bauthor{\bsnm{Brodbeck}, \binits{S.}},
\bauthor{\bsnm{Kamp}, \binits{M.}},
\bauthor{\bsnm{Schneider}, \binits{C.}},
\bauthor{\bsnm{H{\"o}fling}, \binits{S.}},
\bauthor{\bsnm{Yamamoto}, \binits{Y.}}, \betal:
\batitle{Observation of non-{H}ermitian degeneracies in a chaotic
  exciton-polariton billiard}.
\bjtitle{Nature}
\bvolume{526}(\bissue{7574}),
\bfpage{554}--\blpage{558}
(\byear{2015})
\end{barticle}
\endbibitem

\bibitem{zhang2018phonon}
\begin{barticle}
\bauthor{\bsnm{Zhang}, \binits{J.}},
\bauthor{\bsnm{Peng}, \binits{B.}},
\bauthor{\bsnm{{\"O}zdemir}, \binits{{\c{S}}.K.}},
\bauthor{\bsnm{Pichler}, \binits{K.}},
\bauthor{\bsnm{Krimer}, \binits{D.O.}},
\bauthor{\bsnm{Zhao}, \binits{G.}},
\bauthor{\bsnm{Nori}, \binits{F.}},
\bauthor{\bsnm{Liu}, \binits{Y.-x.}},
\bauthor{\bsnm{Rotter}, \binits{S.}},
\bauthor{\bsnm{Yang}, \binits{L.}}:
\batitle{A phonon laser operating at an exceptional point}.
\bjtitle{Nature Photonics}
\bvolume{12}(\bissue{8}),
\bfpage{479}--\blpage{484}
(\byear{2018})
\end{barticle}
\endbibitem

\bibitem{tiwari2013random}
\begin{barticle}
\bauthor{\bsnm{Tiwari}, \binits{A.K.}},
\bauthor{\bsnm{Mujumdar}, \binits{S.}}:
\batitle{Random lasing over gap states from a quasi-one-dimensional amplifying
  periodic-on-average random superlattice}.
\bjtitle{Physical Review Letters}
\bvolume{111}(\bissue{23}),
\bfpage{233903}
(\byear{2013})
\end{barticle}
\endbibitem

\bibitem{wiersig2020review}
\begin{barticle}
\bauthor{\bsnm{Wiersig}, \binits{J.}}:
\batitle{Review of exceptional point-based sensors}.
\bjtitle{Photonics Research}
\bvolume{8}(\bissue{9}),
\bfpage{1457}--\blpage{1467}
(\byear{2020})
\end{barticle}
\endbibitem

\bibitem{pick2017enhanced}
\begin{barticle}
\bauthor{\bsnm{Pick}, \binits{A.}},
\bauthor{\bsnm{Lin}, \binits{Z.}},
\bauthor{\bsnm{Jin}, \binits{W.}},
\bauthor{\bsnm{Rodriguez}, \binits{A.W.}}:
\batitle{Enhanced nonlinear frequency conversion and purcell enhancement at
  exceptional points}.
\bjtitle{Physical Review B}
\bvolume{96}(\bissue{22}),
\bfpage{224303}
(\byear{2017})
\end{barticle}
\endbibitem

\end{thebibliography}
\end{document}